\documentclass[letterpaper, 10 pt, conference]{ieeeconf}  

\IEEEoverridecommandlockouts                              

\usepackage{soul}
\usepackage{color}
\usepackage{comment}
\usepackage{graphicx}
\usepackage{url}
\usepackage{hyperref}

\title{\LARGE \bf
Employing Socially Interactive Agents\\for Robotic Neurorehabilitation Training}

\author{
\authorblockN{Rhythm Arora$^{1}$, Matteo Lavit Nicora$^{2}$, Pooja Prajod$^{3}$, Daniele Panzeri$^{4}$,}
\authorblockN{Elisabeth Andr\'e$^{3}$, Patrick Gebhard$^{1}$, Matteo Malosio$^{2}$}%

\thanks{*This work is funded by the European Union’s Horizon 2020 research and innovation programme under grant agreement No 847926. We thank Charamel GmbH for providing us with the virtual agent technology and the Gloria agent.}%

\thanks{$^{1}$Rhythm Arora and Patrick Gebhard are with the German Research Center for Artificial Intelligence, Saarbrücken, Germany, firstname.lastname@dfki.de.}
        
\thanks{$^{2}$Matteo Lavit Nicora and Matteo Malosio are with the National Research Council of Italy, Institute of Intelligent Industrial Technologies and Systems for Advanced Manufacturing, Lecco, Italy, firstname.lastname@stiima.cnr.it.}%

\thanks{$^{3}$Pooja Prajod and Elisabeth Andr\'e are with the chair for Human-Centered Artificial Intelligence, Institute for Informatics, Augsburg University, Augsburg, Germany, firstname.lastname@uni-a.de.}%

\thanks{$^{4}$Daniele Panzeri is with the Scientific Institute IRCCS E. Medea, Bosisio Parini, Lecco, Italy, firstname.lastname@lanostrafamiglia.it.}}

\begin{document}
\maketitle
\thispagestyle{empty}
\pagestyle{empty}

\begin{abstract}

In today's world, many patients with cognitive impairments and motor dysfunction seek the attention of experts to perform specific conventional therapies to improve their situation. However, due to a lack of neurorehabilitation professionals, patients suffer from severe effects that worsen their condition. In this paper, we present a technological approach for a novel robotic neurorehabilitation training system. It relies on a combination of a rehabilitation device, signal classification methods, supervised machine learning models for training adaptation, training exercises, and socially interactive agents as a user interface. Together with a professional, the system can be trained towards the patient's specific needs. Furthermore, after a training phase, patients are enabled to train independently at home without the assistance of a physical therapist with a socially interactive agent in the role of a coaching assistant.

\end{abstract}

\section{Introduction}

Neurorehabilitation is a widely used medical practice that aims to aid recovery from a nervous system injury. Its purpose is to maximize and maintain the subject's motor control while trying to restore motor functions in people with neurological impairments.

Robot-assisted training has been widely investigated as an effective neurorehabilitation approach that helps augment physical therapy and facilitates motor recovery. According to the literature~\cite{Maciejasz-2014,Zhang-2017,Qassim-2020}, such approaches can help therapists save time and effort while reproducing accurate and repetitive motions and delivering high-intensity training. In particular, upper-limb robotic rehabilitation is one of the fastest-growing areas in modern neurorehabilitation. For this approach, the authors chose to exploit the PlanArm2 prototype~\cite{Yamine20201}, a planar robotic device designed for domestic upper-limb neurorehabilitation.

The system is envisioned to be used by patients at home after an initial calibration phase supervised by professionals. However, a crucial issue for rehabilitation training is user engagement and motivation~\cite{Blank2014184}, which may be lacking if the rehabilitation system is used without a human medical coach. In these terms, we believe that introducing socially interactive agents could represent a valid solution to restore and augment the social aspects typical of an in-person rehabilitation session.

Moreover, inheriting results typical of social robotics, the enhancement of a rehabilitation system through physiological and social signals awareness may augment the system personalization level and further facilitate neuroplasticity. Therefore, the capability of a neurorehabilitation training system to model the patient's state and tune its behavior according to both the measured performance and the inferred social and physiological state could improve the engagement of the user and the outcome of the therapy.

 \begin{figure}[ht]
 \centerline{\includegraphics[width=0.49\textwidth]{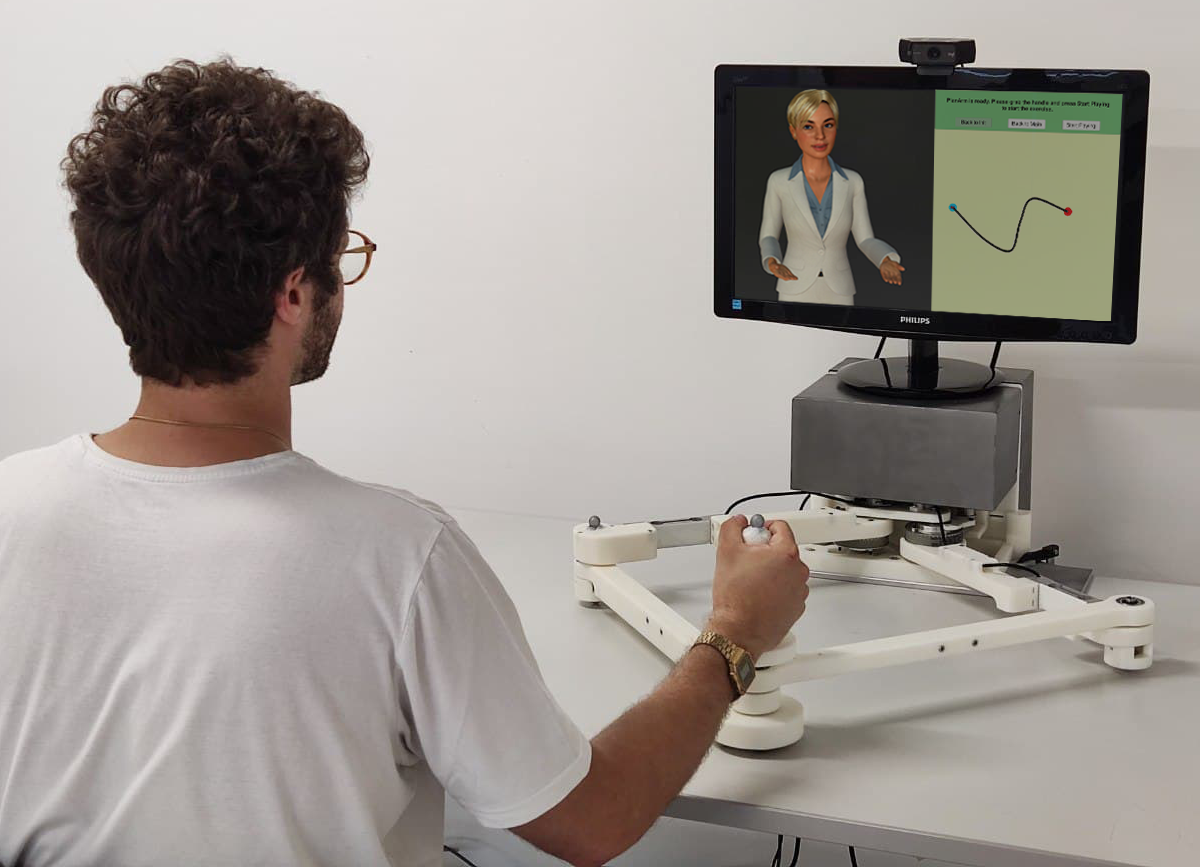}}
 \caption{Empathetic Neurorehabilitation robotic training system concept.}
 \label{fig:system_concept}
 \end{figure}

This paper investigates a technologically-supported upper-limb neurorehabilitation approach by introducing social interaction capabilities in a heterogeneous system composed of 1) a planar rehabilitation device for physical assistance interaction and 2) two socially interactive virtual agents in the role of a coaching assistant, one demanding, one supporting. In this context, physiological and social signals of the participant are analyzed and elaborated into attentiveness, stress, engagement, and pain information, which are then exploited to tune the rehabilitation session coherently.

\section{Background and Related Work}

\subsection{Neurorehabilitation Best Practice}
\label{sec:bpnt}

The constantly growing employment of technological devices in neurorehabilitation therapy can be explained by the introduced ease in reaching a significant number of movement repetitions in a chosen district and in obtaining higher patient engagement, fundamental aspects in the motor rehabilitation process. However, to do so, one must be able to reconcile movement repetition, often leading to boredom, with variability and incentive to enhance the patient's attention and commitment. 
This section reports "neurorehabilitation best practices" we collected from experienced physiotherapists how they manage these issues. Based on these, a system concept is proposed.

During every therapy session, a therapist assesses the actual level of attention, commitment, engagement, and stress or pain currently experienced by the patient to deliver the correct amount of exercise and avoid the risk of too easy or too difficult exercises that may lead to a decay in interest or even frustration. In particular, the therapist monitoring the therapy session relies on the activity scores and the subject's behavior to reach this goal. For example, if the patient cannot achieve a particular performance, the selected activity is likely too difficult. On the other hand, if the patient can perform the exercise but, after some time, becomes very talkative and less performing, it is likely that decay in interest is being experienced. On this basis, the therapist must give feedback, if needed, support when the activities are too difficult or change exercise when the attention starts to decrease.

Furthermore, considering patients affected by neurological impairments, attention problems are frequent. Recording the period to which the attention could last can be useful information to provide a correct dosage of exercise.

Lastly, the management of neurological disorders can be considerably different between adults and children. In both cases, understanding when a pause or a change of exercise is necessary is crucial. However, considering adults, one can count on their responsibility to train towards an improvement, even if the activity could lead to boredom. On the other hand, children may not behave in the same way and, to augment their engagement, it is crucial to introduce gaming aspects to the exercise.

\subsection{Socially Interactive Agents as Medical Coaches}
\label{sec:agents4medicaltraining}

As a use case for interactive social agents, technologically supported health care has been researched for about 15 years. One of the early systems is the Fit Track system with the relational agent Laura \cite{Bickmore-et-al-05}. Laura has the role of an exercise advisor that interacts with patients daily for one month to motivate them to exercise more. A more recent system employs the socially interactive agent Gloria for stress management training system using biofeedback \cite{Schneeberger-et-al-21}. The training system is designed to run autonomously without a medical coach present during the training. The authors conducted an expert interview and a user study in which the novel approach was compared to the stress management training using stress diaries.

The actual training combines two methods of feedback: 1) situation awareness and 2) social interaction feedback with an awareness tool and a socially interactive agent. This approach draws on theories of embodied learning, which proclaims an interaction between bodily and cognitive functions and foresee their concurrent stimulation for more intuitive, long-lasting, and transferable learning \cite{puhl2015blending}. In the study, the awareness tool helped the user to track their sensory-motor reactions to stressful situations, increasing situation (introspective) awareness \cite{price2018interoceptive}. The agent first created a social background to embed this awareness. After a first awareness training, the awareness tool is faded out, and the social agent overtakes the feedback. This implements prominent theories on the socio-emotional routes of learning and implicit emotion regulation in particular \cite{braunstein2017explicit}. In the current context, we are looking into embedding the cognitive understanding of the task in a socio-emotional context to increase the learning effect of the neurorehabilitation training.
The results of a study revealed, HRV biofeedback training supported by a socially interactive agent tends to lower stress levels and improve adaptability to stressful situations.

For the presented neurorehabilitation approach, we rely on the Gloria biofeedback training system. However, the behavior and the training strategies are adapted to the specific needs of neurorehabilitation training.

\subsection{Robotic Neurorehabilitation}

In the last two decades, robotic devices for neurorehabilitation have been widely investigated, developed, and introduced in the market to offer a valid alternative to conventional therapy and fill the constantly growing gap between supply and demand, both for upper limb and lower limb rehabilitation~\cite{Maciejasz-2014,Zhang-2017,Qassim-2020}.

The goal of control algorithms for robotic therapy is to control robotic devices designed for rehabilitation exercises so that the participant's selected exercises provoke motor plasticity and improve motor recovery. Thus, the capability of a robot to adapt the level of assistance according to the skill and the performance of the patient is one of the most crucial features~\cite{Marchal-Crespo-2009, Meng-2015}. Furthermore, assistance is sometimes obtained not only from the biomechanical point of view but also evaluating physiological and psychological aspects~\cite{Novak-2011, Malosio-2016}, for a more fine-grained assessment of the patient's state.

Visual feedback, sometimes including immersive and augmented reality, is an almost indispensable element for robotic rehabilitation devices, to administer therapeutic exercises to increase user engagement, to propose visually-augmented exercises, even resembling Activities of Daily Living (ADLs).
Enriched communication, including virtual coaching, could increase the effect of personalized and motivating communication. This approach is more and more investigated and exploited in medical care, but its exploitation is still limited in the rehabilitation field~\cite{Tropea-2019}.

\begin{figure*}[thpb]
    \centering
    \includegraphics[width=0.65\textwidth, trim=0 38 0 20, clip]{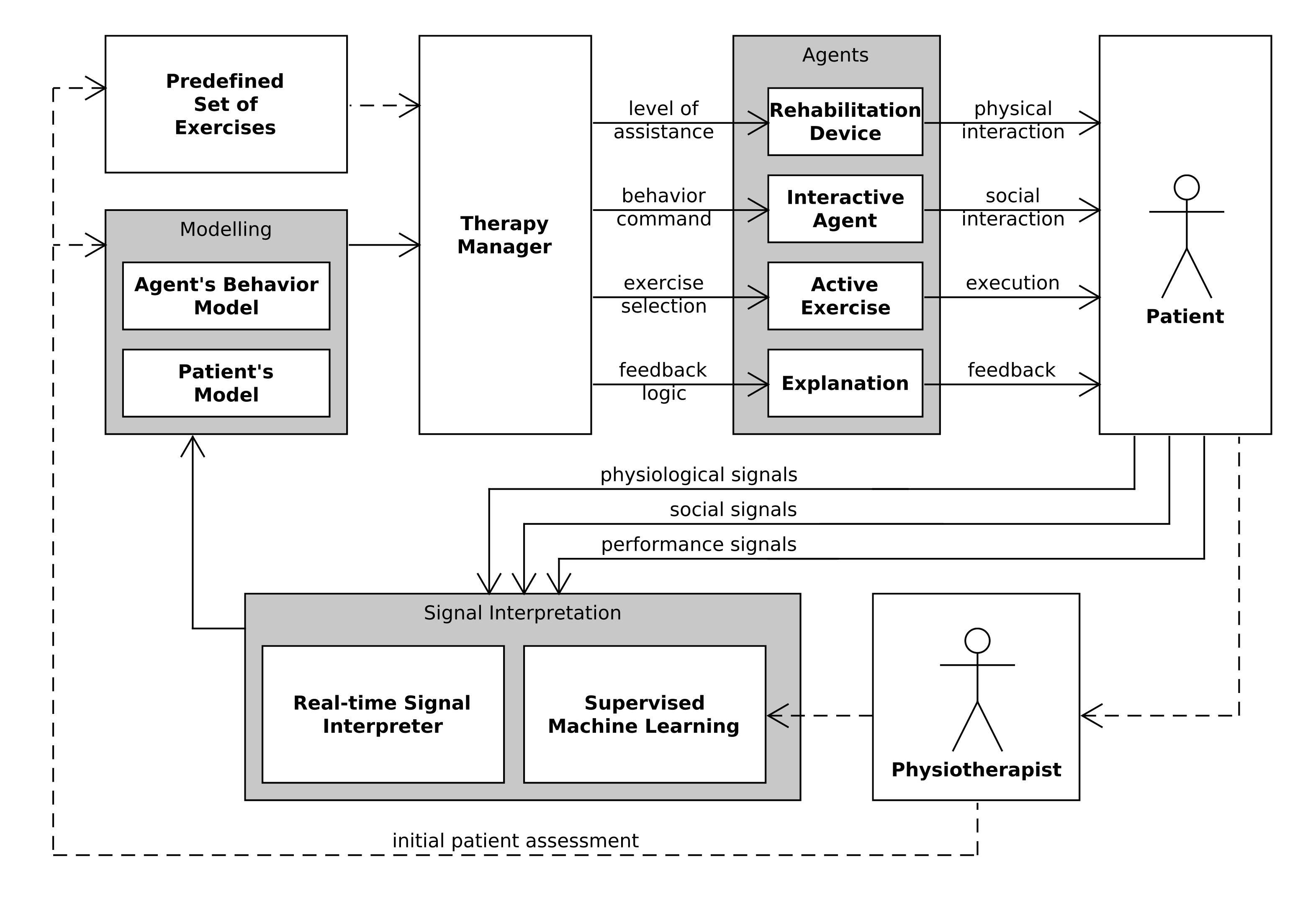}
    \caption{Human-in-the-loop Empathetic Neurorehabilitation Trainer concept}
    \label{fig:deployment}
\end{figure*}

\subsection{Social and Physiological Signals} 

The level of attention a user pays to a task is usually studied in fields such as education (e.g., measuring student attentiveness during a lecture), driver assistance (e.g., monitoring attentiveness of driver to prevent accidents), etc. As noted in \cite{Cramer-2011-short}, attention is a key modulator of neuroplasticity, and information regarding the attentiveness of the patient can be important for the rehabilitation strategy. \cite{smith2003determining} demonstrated that a driver's attention state could be determined through the head and facial features. These features were extracted from a video sequence that was captured through a single camera. \cite{c12} proposed a student attention monitoring system that uses facial and body features extracted from MS Kinect 2D and 3D data. Similar to these studies, we rely on facial and body features to predict the attention level of the patient.

The patient may experience stress while performing the exercise, especially if they find the exercise hard or cannot complete the recommended exercise. Many studies \cite{c7, c8, c9} have studied stress-induced in different scenarios like public speaking, mental arithmetic tests, driving on challenging routes, hyperventilation, etc. These studies found that features extracted from ECG data like Heart Rate Variability (HRV) and EDA data like Skin Conductance Response (SCR) are good stress indicators.

Pain assessment and management are critical for a variety of illnesses and medical interventions. Recent efforts in affective computing suggest that automatic detection of pain from facial expression is a feasible goal. In this work, we adopt a deep learning method to develop a model to detect pain from facial expressions. Typically, this requires a large amount of training data. Still, the availability of pain data is limited due to patient contact, privacy concerns, and the adherence to strict ethical guidelines. However, previous studies like \cite{c6} have demonstrated that a transfer learning approach can be adopted to develop pain recognition models that perform well, despite the limited training data.

\section{Concepts}

The envisioned Empathetic Neurorehabilitation Trainer architecture is reported in Fig.~\ref{fig:deployment}. The goal of the proposed structure is to provide the patient with socially aware neurorehabilitation therapy. For this purpose, the system is equipped with a rehabilitation-specific robotic device and a virtual socially interactive agent (Fig.~\ref{fig:system_concept}). Both device and agent can adapt their behavior based on the patient's performance, as in most assistance-as-needed paradigms, and take into account the patient's physiological and social state. Moreover, these two socially interactive agents, together with the specific task to be carried out and the feedback media chosen to provide the patient with an explanation regarding the exercise, are intended to work as a single entity, actively collaborating to improve the rehabilitation session outcomes further.

All software and hardware components are realized as nodes in a ROS framework \cite{quigley2009ros}. There, a closed-loop monitoring of a set of heterogeneous parameters is introduced. In fact, during the execution of the task, the robotic device is in charge of collecting data regarding the kinematics of the patient's movement (e.g., position, speed), and a wearable device is used to extract a series of physiological values (e.g., ECG, EDA). At the same time, a camera captures the patient's upper body for social signal interpretation purposes. These raw data represent the input for a signal interpretation module (Fig.~\ref{fig:deployment}, left bottom), responsible for providing a series of higher-level quantities such as patient's performance, attentiveness, stress, amusement, and pain.

As depicted (Fig.~\ref{fig:deployment}, right bottom), a human physiotherapist has a central role in the proposed approach. In fact, professional expertise is required for the patient's initial assessment (e.g., residual mobility, attention span), used to define the backbone of a model for both the patient, the agent and a selection of suitable exercises. Moreover, a supervised machine learning module (Fig.~\ref{fig:deployment}, center bottom) is employed to learn from the physiotherapist how to optimally balance the target execution performance for the exercise and the social experience for the specific patient. In fact, a certain level of stress is beneficial for the therapy, and therefore the system is supposed to tune the exercise difficulty accordingly. Also, both challenging and entertaining portions of the session are be included to maximize the patient's attention.

Closing the loop, a Therapy Manager (Fig.~\ref{fig:deployment}, center top) actively exploits the inferred information to decide how the behavior of the socially interactive agent should be changed, which explanations (cf. awareness) should be given, and which exercise and difficulty level should be activated to optimize the therapy experience and effectiveness. The Therapy Manager is a software framework for authoring, orchestrating, and executing scenario content with task specifications. This component will be implemented using the Visual~SceneMaker (VSM) tool \cite{Gebhard-et-al-12}. In addition, VSM provides dedicated plugins for all other components using ROS communication protocol.

\subsection{Interactive Agent-based Training and Explanations}

We employ two socially interactive agents for the proposed agent-based training, one supportive agent and one demanding agent. Both take the role of a coaching assistant. According to the patient's level of training, the agent could be chosen. 
The supportive agent aims to enhance the patient's motivation by creating a warm, friendly, and motivating atmosphere during the training sessions. The agent signals comprehension by head nods and brief verbal utterances, such as" OK" to create a pleasant atmosphere for the patient. The patient is also encouraged through the use of positive feedback, such as smiles \cite{Gebhard-et-al-14}. On the other hand, we have a demanding agent that appears when the patient is seen to be losing attention or interest (Sec.~\ref{sec:ssi}). Initially, the patient starts the training session with a good performance, but their performance starts deteriorating over time. When such a situation occurs, the demanding agent appears, where the patient is provided with activities with a higher level of complexity. There, the demanding agent exhibits a gaze behavior that is supposed to be perceived as dominant and shows fewer head tilts while speaking and listening \cite{10.1007/978-3-540-85483-8_19}.

Based on the described concept of assessing the affective states such as attentiveness, stress, and pain (Sec.~\ref{sec:ssi}), the agent adapts its motivational strategy. Here, we are considering the possibility of having two agents. One agent supports the patient in the case of actual difficulty during the training. The other shows a more demanding behavior and can be used if the patient is losing attention or interest, which can be detected with the help of social and physiological signals. If the exchange of them during the training exercises supports the training regarding, for example, efficiency, it is part of future investigations.

The coaching assistant follows the best practices of training professionals (Sec.~\ref{sec:bpnt}). The training always starts with a pre-screening phase. This phase serves for the identification of patient issues and the selection of further training sessions. In the pre-screening session, the physiotherapist observes how the patient interacts with the agent and the device and what the patient can achieve. In addition, the movement space is personalized, e.g., space is increased if a noticeable improvement in the patient's movement is detected. Since every patient is different from the other, they might face difficulties in different areas, e.g., some patients may find it difficult to control the smoothness of movement. Others may have difficulties in gripping.

The exercises selected for all patient resembles actual day-to-day useful movements. During the training sessions, time and difficulty are tailored automatically. The system adapts to the level of attention or stress. For example, if the level of attention of the patient decreases, more competitive or challenging exercises/tasks cannot be provided. Strategies to overcome this phase to suggest a training pause or offer playful intermezzos. The particular realization always follows best practices (Sec.~\ref{sec:bpnt}), and known approaches (Sec.~\ref{sec:agents4medicaltraining}) are part of future investigations.

Content-wise, each session is represented by small training fragments. By combining such fragments, the difficulty and time of a training session can be tuned. In addition, each session is followed by some relaxation period, which could involve playful, less stressful activities.

\subsection{Neurorehabilitation Training Device}

This work exploits the PlanArm2 rehabilitation device~\cite{Yamine20201}, a 2-DOF planar robotic prototype validated for its active, passive, and assistive upper-limb rehabilitation capabilities.

The selected device has been developed specifically to achieve an acceptable compromise regarding workspace symmetry concerning the sagittal plane, relatively large workspace, portability, and affordability, rendering it suitable for domestic training sessions. Furthermore, from a long-term point of view, the system could be autonomously used by patients directly at home after an initial calibration phase supervised by professionals. In these terms, the presence of a socially interactive agent could provide the missing social interaction and improve the patient's experience while still guaranteeing an effective training session. However, not all cases are suited for autonomous rehabilitation, which stresses the importance of the patient's initial assessment performed by a specialist.

Besides the agent, the device itself is supposed to adapt in a social way. Therefore, the envisioned assistive control logic exposes a series of parameters intended to be automatically tuned not only based on the patient's performance, as in most assist-as-needed paradigms, but also considering the patient's physiological and social inferred state.

An example of an adjustable parameter is the target speed of execution. By increasing this value, different effects can be expected: in some cases, the exercise may become too stressful and might generate frustration or pain in the patient. In other cases, it could render the activity more challenging and augment the patient's engagement. By monitoring performance, social and physiological signals, the system should be able to tune the parameter correctly, also based on the therapist's experience embedded in the Supervised Machine Learning (Fig.~\ref{fig:deployment}, center bottom) module.

Similarly, a series of additional parameters are available for adaption in the context of the proposed assistive rehabilitation controller. For instance, given an ideal target trajectory to be followed by the patient, the level of regular and tangential assistance force provided can be independently tuned to the patient's needs. By regulating these assistance parameters, one could range from entirely passive to completely active rehabilitation paradigms, meaning that the same device could be used for a wide range of patients after a therapist has properly tuned and trained the system.

\subsection{Signal Interpretation}
\label{sec:ssi}

During the training sessions, the social signals from the patients can be used to infer useful information about their experience. We identified the following emotional states that would serve as an input to the agent's motivational strategy.

\subsubsection{Attentiveness} The patient may get distracted during the training session because of boredom, lack of motivation. Motivation and attention can be critical modulators of neuroplasticity, contributing to determining the actual outcome of a rehabilitation therapy~\cite{Cramer-2011-short}. Therefore, the level of attention of the patient is an important input to the agent's motivational strategy in neurorehabilitation. Previous works have demonstrated that attentiveness can be predicted through physiological signals like EEG \cite{c11} and facial and body pose features like gaze direction, head orientation, body posture \cite{c12}. Deep learning models are trained to focus on facial and body features that can be extracted using the camera.

\subsubsection{Stress and amusement} These affective states are crucial as we introduce some elements of gamification to the training session. On the one hand, the patient may be stressed, for example, when the exercise is too hard. On the other hand, the patient may enjoy the gamification and experience flow. The agent's behavior would differ depending on which state is detected. Physiological signals such as ECG and EDA have been demonstrated to be very effective in detecting stress \cite{c7, c8}. The WESAD data set \cite{c9} is suitable for our use case as it contains multiple physiological signals from wearable sensors and is annotated for the states we intent to classify (stress, amusement, neutral). We adapt the neural network architecture proposed in \cite{c10} to train an LSTM (Long short-term memory) network to detect stress and amusement.

\subsubsection{Pain} Pain is an important social component, as the expression of pain triggers social reactions such as empathy and care \cite{c2}. Many health-care related fields are deploying image or video based automatic pain detection \cite{c3, c4}. Our idea is to detect pain from the facial expressions of the patient captured by the front camera. To achieve this, we train a deep learning model that can discern pain and no-pain images. We use the images from UNBC-McMaster shoulder pain expression database \cite{c5} to train our model. This data set contains images of 25 participants suffering from shoulder pain while performing a range of motions with their affected and unaffected limb - a scenario similar to our use case. The small number of unique samples in this data set would affect the performance of our model. Similar to \cite{c6}, we use a transfer learning approach to tackle this.

To ensure that our models are generalizable and not over-tuned on a specific dataset, we perform cross-dataset validations by testing our model on different datasets.

\subsection{Socially Interactive Agent and Visual Explanations}

For the presented approach, a socially interactive agent together with the current training task and explanations will be displayed on a monitor Fig.~\ref{fig:system_concept}). The agent acts as a mediator to motivate, inform, and help the patient carry out specific neurorehabilitation tasks. The agent's behavior and the rehabilitation device are tuned in a way that it seems like they are one entity, and the agent is helping the patient apply a certain amount of force. Another task of the agent is to verbally give explanations of the current (and past) states of the training if asked by the patient or if best practice strategies (Sec.~\ref{sec:bpnt}) suggest giving explanations to overcome specific training situations pro-actively.

Technology-wise, the agent's behavior is modeled with the open-source Visual~SceneMaker (VSM, \href{http://scenemaker.dfki.de}{scenemaker.dfki.de}) toolkit \cite{Gebhard-et-al-12}. The tool is explicitly designed to facilitate studies with interactive agents. It comes with a real-time execution and authoring component for modeling verbal and non-verbal behavior of virtual agents and system actions. VSM also allows adapting an agent's behavior relying on external information procedurally. The agent's contributions and behavior are influenced by the patient's social and physiological signals, and these signals are interpreted in real-time by a social signal interpretation framework (SSI, \href{http://openssi.net}{openssi.net}) (Sec.~\ref{sec:ssi}).

The Gloria agent is a high-quality agent with a natural human appearance and verbal and nonverbal dialogue skills. The agent software independently runs in a web browser (\href{https://www.charamel.com}{charamel.com}). The agent is capable of performing social cue-based interaction with the user. It performs lip-sync speech output using the state-of-the-art CereProc (\href{https://www.cereporc.com}{cereproc.com}) Text-To-Speech system. For a more advanced animation control, Gloria allows the direct manipulation of skeleton model joints (e.g.,  neck joint). In addition, it comes with 36 conversational motion-captured gestures, which can be modified during run-time in some aspects (e.g., overall speed). Moreover, it provides 14 facial expressions, which contains, among others, the six basic emotional expressions defined by Ekman \cite{Ekman92}.

\section{Summary and Future Work}

In this paper, we have presented an approach for an empathetic robotic neurorehabilitation system to help the patients train at home without a therapist but with a socially interactive agent in the role of a coaching assistant. To ensure proper and suitable training exercises, the system is prepared for patients with the help of a professional physiotherapist before it can be used at home. Our vision is to maximize the training effect by measuring social signals, e.g., attentiveness and stress, and physiological signals, e.g., EEG and EDA, to tailor the time and difficulty of training exercises. The interactive social agent motivates and gives explanations. Patients can choose if the agent should behave supportive or demanding.

The following steps consist of evaluating several aspects concerning training efficiency with objective measurements (e.g., overall speed, precision) and standard questionnaires. Furthermore, we investigate if the presence of an interactive agent during therapy sessions has a significant impact on the user's engagement and training progress. Finally, another future evaluation asks if the social and explanatory skills of the system (with or without an agent) impact the training progress?




\bibliographystyle{Bibliography/IEEEtran}
\bibliography{Bibliography/IEEEabrv,Bibliography/bibfile}

\addtolength{\textheight}{-12cm}   

\end{document}